\documentclass[12pt]{JHEP3}

\usepackage{epsfig}

\newcommand {\chin} [1] {\tilde{\chi}^0_{#1}}
\newcommand {\chm} [1] {\tilde{\chi}^-_{#1}}
\newcommand {\chp} [1] {\tilde{\chi}^+_{#1}}
\newcommand {\chpm} [1] {\tilde{\chi}^\pm_{#1}}
\newcommand {\cost} {\cos \theta_{\tilde t}}
\newcommand {\eq} [1] {Eq.~(\ref{#1})}
\newcommand {\fig} [1] {Fig.~\ref{#1}}
\newcommand {\figs} [2] {Figs.~\ref{#1} and \ref{#2}}
\newcommand {\mchin} [1] {m_{\tilde{\chi}^0_{#1}}}
\newcommand {\mchpm} [1] {m_{\tilde{\chi}^\pm_{#1}}}
\newcommand {\mg} {m_{\tilde g}}
\newcommand {\mst} {m_{{\tilde t}_1}}
\newcommand {\sect} [1] {Section~\ref{#1}}
\newcommand {\tab} [1] {Table~\ref{#1}}

\title{The decays $\tilde g \to \tilde t_1 \, \bar{b} \, W^-$ and  
        $\tilde g \to \tilde t_1 \bar{c}$ and  phenomenological
       implications in supersymmetric theories}
\author{W.~Porod \\
        Inst.~f\"ur Theor.~Physik, Universit\"at Z\"urich, CH-8057 Z\"urich,
       Switzerland \\ E-mail: \email{porod@physik.unizh.ch}
      }

\abstract{  
  We show that the decay $\tilde g \to \tilde t_1 \, \bar{b} \, W^-$
  is important and can even be dominant in the region of parameter space
  where it is kinematically allowed. We discuss phenomenological
  implications within the Minimal Supersymmetric Standard Model and
  models with broken R-parity. We consider the flavour diagonal case
  as well as a possible mixing between squarks of different
  generations. In the latter case also the decay $\tilde g \to \tilde
  t_1 \bar{c}$ is potentially important.  We show that the decay 
  $\tilde g \to \tilde t_1 \, \bar{b} \, W^-$ is
  sensitive to the stop mixing angle. Furthermore we demonstrate that
  in scenarios with a higgsino--like LSP the gluino decays mainly into
  final states containing top quarks or a light stop if allowed by
  kinematics.}

\preprint{ZH-TH 6/02}

\begin{document}

\section{Introduction}
\label{sec:intro}

At the Tevatron as well as at the future Large Hadron Collider (LHC) the
search for supersymmetric particles is among the main topics of their
experimental programs. Here the strongly interacting supersymmetric
partners of quarks and gluons, squarks and gluinos, are expected
to have the largest cross sections. Their production as well as their
decays have therefore been intensively studied in recent years
\cite{Krutelyov:2000ia,GluDecaysOld,Abel:2000vs}.

In these studies it has been assumed that the gluino $\tilde g$ decays either
into $q \tilde q_i$ if kinematically allowed or into
$q \bar{q} \chin{i}$, $q' \bar{q} \chpm{j}$ and $g \chin{i}$ otherwise.
Here $\tilde q_i$, $\chin{i}$, $\chpm{j}$ denote squarks, neutralinos and 
charginos, respectively.
These decay modes have been used in the searches for gluinos
(see e.g.~\cite{experiment} and references therein). However, there
exists also the possibility that the gluino decays via a three body
decay into the lighter stop, namely: 
$\tilde g \to \tilde t_1 \, \bar{b} \, W^-$. The necessary mass
hierarchy $m_{\tilde q}, m_{{\tilde t}_1} + m_t > m_{\tilde g}
> m_{{\tilde t}_1} + m_W + m_b$ can be obtained e.g.~in the minimal
supergravity model as will be shown below. 
 In the case that this mass
hierarchy is realized in nature there are further
gluino decays, violating flavour, into $\tilde t_1 \bar{c}$.
To our knowledge this interesting possibilities have not been treated
so far in the literature. 

In this paper we  discuss the decays 
$\tilde g \to \tilde t_1 \, \bar{b} \, W^-$ and $\tilde t_1 \bar{c}$
and possible signatures. Our framework is  mainly the Minimal
Supersymmetric Standard Model (MSSM) \cite{Haber:1984rc} with
conserved R-parity. However, we also discuss possible implications
of R-parity violation for the signatures of these two decay modes.
The paper is organized as follows: in the next section we present
our conventions as well as the formulas for the gluino decays.
In \sect{sec:num} we present our numerical results for various
scenarios, with and without R-parity. Finally in \sect{sec:con} we 
present our conclusions.

\section{Conventions and the formulas for the widths}
\label{sec:formula}

The parameters relevant to our discussions
are the soft susy breaking mass parameters for the squarks $M_{Q_i}$,
$M_{U_i}$, $M_{D_i}$, the trilinear parameters $A_{t,b}$, the
gaugino mass parameters $M_{1,2}$, the gluino mass $\mg$,
the higgsino mass parameter $\mu$ and the
ratio of the Higgs vacuum expectation values $\tan \beta = v_2/v_1$.
Here $i$ is a generation index.

We  give the formulas for complex parameters to be as general as 
possible although later in the numerical discussions we confine ourselves
to real parameters to reduce the numbers of free parameters. 
It is well known that in the third generation left and right squarks
mix due to the presence of the large Yukawa couplings. 
We give here for completeness the
formulas for the mass matrix matrices as well as for the mass-eigenstates.
Neglecting  a possible generation mixing the mass matrices read as:
\begin{eqnarray}
 M^2_{\tilde q} = \left( \begin{array}{cc}
      M^2_{\tilde q,LL} & M^2_{\tilde q,RL} \\
      M^2_{\tilde q,LR} & M^2_{\tilde q,RR} \end{array} \right) \hspace{3cm}
    (q=b,t)
\end{eqnarray}
with 
\begin{eqnarray}
M^2_{\tilde t,LL} &=& M^2_{Q,3} + m^2_t
                  + (\textstyle \frac{1}{2} - \frac{2}{3} \sin^2 \theta_W )
                      \cos 2 \beta \, m^2_Z \\
M^2_{\tilde t,LR} &=& \left(M^2_{\tilde t,RL} \right)^*
           = m_t (A_t - \mu^* \cot \beta ) \\
M^2_{\tilde t,RR} &=& M^2_{U,3} + m^2_t
          + \textstyle \frac{2}{3} \sin^2 \theta_W  \cos 2 \beta \, m^2_Z \\
M^2_{\tilde b,LL} &=& M^2_{Q,3} + m^2_b
                  - (\textstyle \frac{1}{2} - \frac{1}{3} \sin^2 \theta_W )
                      \cos 2 \beta \, m^2_Z \\
M^2_{\tilde t,LR} &=& \left(M^2_{\tilde t,RL} \right)^*
           = m_b (A_b - \mu^* \tan \beta ) \\
M^2_{\tilde t,RR} &=& M^2_{D,3} + m^2_b
              - \textstyle \frac{1}{3} \sin^2 \theta_W  \cos 2 \beta \, m^2_Z
\end{eqnarray}
The mass eigenstates $\tilde q_i$  are $(\tilde q_1, \tilde q_2)=
(\tilde q_L, \tilde q_R) {\mathcal{R}^{\tilde q}}^T$ with
 \begin{equation}
\mathcal{R}^{\tilde q}=\left( \begin{array}{ccc}
e^{i\varphi_{\tilde q}}\cos\theta_{\tilde q} & 
\sin\theta_{\tilde q}\\[5mm]
-\sin\theta_{\tilde q} & 
e^{-i\varphi_{\tilde q}}\cos\theta_{\tilde q}
\end{array}\right),
\label{eq:rtau}
\end{equation}
\begin{eqnarray}
\cos\theta_{\tilde q}&=&\frac{-|M_{\tilde q,LR}^2|}
   {\sqrt{|M_{\tilde q, LR}^2|^2+(m_{\tilde q_1}^2-M_{\tilde q,LL}^2)^2}}
 \, \, , \, \, 
\sin\theta_{\tilde q}=\frac{M_{\tilde q,LL}^2 - m_{\tilde q_1}^2}
   {\sqrt{|M_{\tilde q, LR}^2|^2+(m_{\tilde q_1}^2-M_{\tilde q,LL}^2)^2}}
 \, \, , \nonumber \\
\varphi_{\tilde q} &=& \mathrm{arg}(M_{\tilde q, LR}) \, .
\label{eq:thtau}
\end{eqnarray}
The mass eigenvalues are
\begin{equation}
 m_{\tilde q_{1,2}}^2 = \frac{1}{2}
 \left( (M_{\tilde q,LL}^2+M_{\tilde q,RR}^2) \mp 
\sqrt{(M_{\tilde q,LL}^2 - M_{\tilde q,RR}^2)^2 +4|M_{\tilde q,LR}^2|^2}
\right).
\label{eq:m12}
\end{equation}

\begin{figure}[t]
\setlength{\unitlength}{1mm}
\begin{center}
\begin{picture}(83,30)
\put(-70,-120){\mbox{
    \epsfig{figure=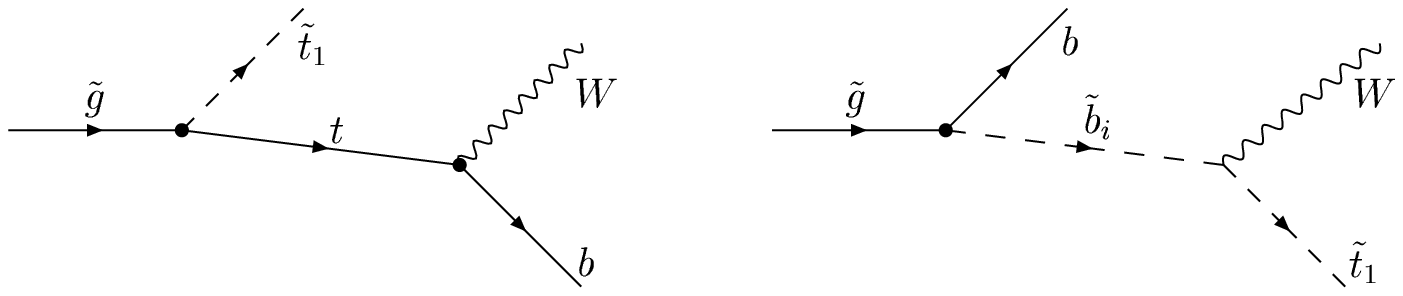,height=28.cm,width=19.cm}}}
\end{picture}
\end{center}
\caption[]{Feynman graphs for the decay
     $\tilde g \to \tilde t_1 \, \bar{b} \, W^-$.}
\label{fig:graphs}
\end{figure}

The decay $\tilde g \to \tilde t_1 \, \bar{b} \, W^-$ proceeds via
top--quark exchange and sbottom exchange as shown in \fig{fig:graphs}.
The relevant part of the Lagrangian is given by:
\begin{eqnarray}
  \label{eq:lagrangian}
  {\cal L} &=&
   \lambda^a_{rs} \bar b_r \left( c^{\tilde b_i}_L P_L
                 + c^{\tilde b_i}_R P_R \right) {\tilde g}^a \tilde b_{i,s}
  + \lambda^a_{rs} \bar t_r \left( c^{\tilde t_i}_L P_L
                + c^{\tilde t_i}_R P_R \right) {\tilde g}^a \tilde t_{i,s}
 \nonumber \\ &+& C_{tb} W^+_\mu \bar{t} \gamma^\mu b 
       + d_{ij} W^+_\mu \left( i \tilde b_i \partial^\mu \tilde{t}^*_j
                             - i \tilde{t}^*_j  \partial^\mu \tilde b_i
                        \right) + h.c.
\end{eqnarray}
Here $r,s$ are color indices whereas $i,j$ denote the mass eigenstates.
The $\lambda^a_{rs}$ are the Gell--Mann matrices with normalization
$\sum_a Tr(\lambda^a)^2 = 16$.
The couplings are given by:
\begin{eqnarray}
  \label{eq:coupl}
  \begin{array}{llr}
  c^{\tilde q_i}_L = \frac{g_3 R^{\tilde q}_{i2} e^{- i \phi_3/2}}{\sqrt{2}}
  &
  , \, \, 
   c^{\tilde q_i}_R = -\frac{g_3 R^{\tilde q}_{i1} e^{i \phi_3/2}}{\sqrt{2}}
  & (q=b,t) 
  \\
  d_{ij} = \frac{g}{\sqrt{2}} \left( R^{\tilde t}_{j1} \right)^*
                                 R^{\tilde b}_{i1} V_{tb}
  & , \, \, C_{tb} = - \frac{g}{\sqrt{2}} V_{tb}
 & \\
 \end{array}
\end{eqnarray}
Here $\phi_3$ is the phase of the gluino mass parameter $M_3$ and
$V_{tb}$ is the (33) element of the CKM matrix.
The partial width can be written as
\begin{eqnarray}
  \label{eq:width}
  256 \pi^3 m^2_{\tilde g} \frac{d \Gamma}{d s d t} &=& 
 \frac{|C_{tb}|^2}{(s-m_t^2)^2 + m_t^2 \Gamma_t^2} \nonumber \\
  && \times \Bigg[ |c^{\tilde t_1}_R|^2 
              \bigg(
               \left(m^2_{\tilde g} - m_{{\tilde t}_1}^2 \right)
               \left( m^2_b + m^2_{\tilde g} -t \right)  
  \nonumber \\  && \hspace{15mm} 
        + \left(m_{{\tilde t}_1}^2 -m^2_{\tilde g}  - s \right)
         \left( u -m_{{\tilde t}_1}^2 - m^2_b \right) 
  \nonumber \\  && \hspace{15mm}  + \frac{1}{m^2_W}
       \left( m_{{\tilde t}_1}^2-m^2_{\tilde g} - s \right)
       \left( s -m^2_b - m^2_W \right)
       \left( t -m_{{\tilde t}_1}^2 - m^2_W \right) 
  \nonumber \\  && \hspace{15mm} 
       + \frac{m^2_{\tilde g} - m_{{\tilde t}_1}^2}{m^2_W}
           \left( s -m^2_b - m^2_W \right)
           \left(m^2_{\tilde g} + m_W^2 - u \right) \bigg)
  \nonumber \\  && \hspace{3mm} 
     + 2 \mathrm{Re}\left( c^{\tilde t_1}_R \left(c^{\tilde t_1}_L\right)^*
           \right) 
       m_{\tilde g} m_t \left( s -2 m^2_W + m^2_b +
          \frac{(s-m^2_b)^2}{m^2_W} \right)
  \nonumber \\  && \hspace{3mm} 
    + |c^{\tilde t_1}_L|^2 \bigg(
        m^2_t \left( m^2_b + m^2_{\tilde g} -t \right)
  \nonumber \\  && \hspace{15mm}  
        + \frac{m^2_t}{m^2_W}\left( s -m^2_b - m^2_W \right)
           \left(m^2_{\tilde g} + m_W^2 - u \right)
         \bigg) \Bigg] 
\nonumber \\
 &+& \sum_{j=1}^2 \mathrm{Re} \Bigg[
     \frac{d_{j1} C_{tb}}{(s-m^2_t + i m_t \Gamma_t)
                   (t-m^2_{\tilde b_j} - i m_{\tilde b_j} \Gamma_{\tilde b_j})}
\nonumber \\ && \hspace{9mm} \times \Bigg(
   c^{\tilde b_j *}_R c^{\tilde t_1}_R
   \bigg( 2 m^2_{\tilde g}\left( m^2_b + m^2_{\tilde g} -t \right) 
        - 2 m_{{\tilde t}_1}^2 \left( u -m_{{\tilde t}_1}^2 - m^2_b \right)
\nonumber \\ && \hspace{17mm} + \frac{t -m_{{\tilde t}_1}^2 - m^2_W}{m^2_W}
   \left( 2 s t - m^2_{\tilde g} (s +  m^2_W - m^2_b)
        - 2 m^2_b m_{{\tilde t}_1}^2 \right) \bigg)
\nonumber \\ && \hspace{11mm}  +
   c^{\tilde b_j *}_R c^{\tilde t_1}_L
  \bigg( 2 m_{\tilde g} m_t\left( m^2_b + m^2_{\tilde g} -t \right)
\nonumber \\ && \hspace{26mm}  +
        \frac{m_{\tilde g} m_t}{m^2_W}
        \left( s -m^2_b - m^2_W \right)
             \left( t -m_{{\tilde t}_1}^2 - m^2_W \right) \bigg)
\nonumber \\ && \hspace{11mm}  +
   c^{\tilde b_j *}_L c^{\tilde t_1}_R
  \bigg( 2 m_{\tilde g} m_b
          \left( m_{{\tilde t}_1}^2-m^2_{\tilde g} - s \right)
\nonumber \\ && \hspace{26mm}  -
        \frac{m_{\tilde g} m_b}{m^2_W}
        \left( s -m^2_b + m^2_W \right)
             \left( t -m_{{\tilde t}_1}^2 - m^2_W \right) \bigg)
\nonumber \\ && \hspace{11mm}  +
   c^{\tilde b_j *}_L c^{\tilde t_1}_L
  \bigg( 2 m_t m_b
          \left( m_{{\tilde t}_1}^2+m^2_{\tilde g} - s \right)
\nonumber \\ && \hspace{26mm}  -
        \frac{m_t m_b}{m^2_W}
        \left( t -m_{{\tilde t}_1}^2 - m^2_W \right)
        \left(m^2_{\tilde g} + m_W^2 - u \right)      \bigg) \Bigg) \Bigg]
\nonumber \\
&+&
  \sum_{i,j=1}^2
     \bigg[ \left( (c^{\tilde b_i}_R)^* c^{\tilde b_j}_R 
            + (c^{\tilde b_i}_L)^* c^{\tilde b_j}_L \right)
              (m^2_b + m^2_{\tilde g} -t )
\nonumber \\ && \hspace{11mm}  
           + \left( (c^{\tilde b_i}_R)^* c^{\tilde b_j}_L 
             + (c^{\tilde b_i}_L)^* c^{\tilde b_j}_R \right) 2 m_b m_{\tilde g}
           \bigg]
  \nonumber \\ && \hspace{8mm} \times
    \frac{ d^*_{i1} d_{j1} (2 t + 2 m^2_{\tilde t_1} - m^2_W)}
           {(t-m^2_{\tilde b_i} + i m_{\tilde b_i} \Gamma_{\tilde b_i})
           (t-m^2_{\tilde b_j} - i m_{\tilde b_j} \Gamma_{\tilde b_j})} 
\end{eqnarray}
with $\kappa(x,y,z) = \sqrt{(x-y-z)^2-4 y z}$,
 $s=(p_{\tilde g} - p_{{\tilde t}_1})^2$, 
 $t=(p_{\tilde g} - p_b)^2$ and  $u=(p_{\tilde g} - p_W)^2$.
The total width is obtained by integrating in the range
\begin{eqnarray}
s_{min} &=& (m_W + m_b)^2\, \, , \hspace{4mm} 
s_{max} = (m_{\tilde g} - m_{{\tilde t}_1})^2 \, \, ,\\
t_{min,max} &=& 
  \frac{m_{\tilde g}^2 + m_{{\tilde t}_1}^2 + m_W^2 + m_b^2 - s}{2}
 + \frac{(m_{\tilde g}^2 - m_{{\tilde t}_1}^2)(m_W^2 - m_b^2)}{2s}
 \nonumber \\ &&
 \mp \frac{\kappa(s,m_{\tilde g}^2,m_{{\tilde t}_1}^2)
           \kappa(s,m_W^2,m_b^2)}{2s}
\end{eqnarray}

We will also consider the possibility of $\tilde g \to \tilde t_1 \bar{c}$.
Using the approximate formulas for scalar top -- scalar charm
mixing as given in \cite{Hikasa:1987db} we get: 
\begin{eqnarray}
  \label{eq:gamGStC}
  \Gamma(\tilde g \to \tilde t_1 \bar{c}) =
  \frac{\alpha_s}{8} |\epsilon|^2 m_{\tilde g}
    \left( 1 - \frac{m^2_{{\tilde t}_1}}{m^2_{\tilde g}} \right)
\end{eqnarray}
with
\begin{eqnarray}
\epsilon &=& \frac{\Delta_L R^{\tilde t}_{11} + \Delta_R R^{\tilde t}_{21}}
                {m^2_{{\tilde c}_L} - m^2_{{\tilde t}_1}} \\
\Delta_L & = & - \frac{g^2}{16 \pi^2} \ln \left(\frac{M_X^2}{m_W^2} \right)
\frac{V^{\ast}_{tb} V_{cb} m_b^2 }{2 m_W^2 \cos^2 \beta }
( M_{Q,2}^2 + M_{D,3}^2 + M_{H_1}^{2} + A_b^2 )
\label{deltal} \\[0.2cm]
\Delta_R & = & \frac{g^2}{16 \pi^2} \ln \left(\frac{M_X^2}{m_W^2}
\right)
\frac{V^{\ast}_{tb} V_{cb} m_b^2 }{2 m_W^2 \cos^2 \beta } m_t A_b
\label{deltar}
\end{eqnarray}
where $M_X$ is a high scale which we assume to be the Planck mass to get a
maximal mixing. 
$ M_{H_1} $ is the soft susy breaking  Higgs mass term
and $V_{tb} $ and $V_{cb} $ are the respective elements of the
CKM matrix. Assuming proper electroweak symmetry breaking one 
gets at tree level 
$m^2_{H_1} = m^2_{A^0} \sin^2 \beta - \cos 2 \beta m^2_Z / 2 - |\mu|^2$,
where $m_{A^0}$ is the mass of the pseudoscalar Higgs boson. The formulas
for $\Delta_L$ and $\Delta_R$ are the result of a single step integration
of the corresponding RGEs assuming that the CKM matrix is the only source
of flavour violation at the GUT scale. In this
approximation the parameters $M_{Q,2}^2$,  $M_{D,3}^2$, $M_{H_1}^{2}$, and
$A_b$ can be evaluated at any scale because the induced error would be
of higher orders. Therefore the expression should be treated as a rough
estimate giving the order of magnitude for the mixing. For definiteness
we take the corresponding values of the parameters at the electroweak scale.
In addition one should note that this approximation is an expansion in
$m_b /( m_W \cos \beta)$. Therefore one expects, that for small 
$\tan \beta$ the quality of this approximation is better than for large
$\tan \beta$.

In principle there is also the possibility that the gluino decays according to
$\tilde g \to \tilde t_1 \bar{u}$. However, this decay
is suppressed by $| V_{ub}/ V_{cb}|^2 \simeq 10^{-2}$ 
in the approximation used above
and will therefore be neglected in the following.

Formulas for the decays $\tilde g \to q \bar{q} \chin{i}$, 
 $\tilde g \to q' \bar{q} \chpm{k}$ and $\tilde g \to g  \chin{i}$
can be found in \cite{GluDecaysOld}. Formulas
for the cross section for gluino pair production as well
as associated gluino-squark production including QCD
corrections are given in  \cite{Beenakker:1996ch} and for associated
gluino-gaugino
production including QCD corrections in \cite{Berger:2000iu}.

\section{Numerical Results}
\label{sec:num}

In this section we present our numerical results for three different
frameworks: (i) The MSSM without flavour violation. (ii) The MSSM
with minimal flavour violation. 
In this case we assume that
at an high energy scale the only source of flavour violation is given
by the CKM matrix. RGE running induces in this case non-vanishing
flavour violating couplings couplings between squarks, quarks and gluino. 
However, these couplings have to be small to respect the bounds
on flavour changing neutral currents FCNCs \cite{Gabbiani:1996hi}. (iii) 
A scenario where R-parity is broken by bilinear terms \cite{Diaz:1997xc}.
This class of models resemble in many respects
also the case of R-parity violation
with trilinear terms \cite{Dreiner:wm} violating lepton number as will be
shown below.

The parameter space relevant for our discussion is given
by: $m_{\tilde q}, m_{{\tilde t}_1} + m_t > m_{\tilde g}
    > m_{{\tilde t}_1} + m_W + m_b$. In the following we take
the parameters freely without referring to a high scale scheme such
as minimal supergravity (mSUGRA) or gauge mediated supersymmetry breaking
(GMSB). However,
we want to stress that this mass hierarchy can be obtained in mSUGRA.
This can be seen by plugging the following 
approximate solutions of the 1-loop RGEs
(see e.g.~\cite{Bartl:2001wc} and references therein) in the inequalities
above:
$m_{\tilde q}^2 \simeq M^2_0 + 6.2 M^2_{1/2}$, 
$m_{\tilde g} \simeq 3.5 M_{1/2}$ and 
$m^2_{{\tilde t}_1} = 0.43 M^2_0 + 4.55 M^2_{1/2} + m^2_t + 0.2 M_{1/2} A_0
 - M_{1/2} \sqrt{2.25 M^2_{1/2}+1.13 M^2_0 +20.2 m^2_t}/2$
where $M_{1/2}$, $M_0$ and $A_0$ are the universal gaugino mass parameter,
the universal scalar mass and the universal trilinear coupling at the GUT
scale. 

In all numerical examples below we take $\mg = 500$~GeV,
$m_{{\tilde t}_2} = 660$~GeV and for the
first two generation squark mass parameters 600 GeV. Moreover, we fix
$A_b = - 865$~GeV.  The relevant Standard Model parameters used are:
$m_t = 174.3$~GeV, $m_b = 4.6$~GeV, $\sin^2 \theta_W = 0.2315$,
$\alpha_s(m_Z) = 0.118$ and $\alpha(m_Z) = 1 / 127.9$.  The parameters
$m_{{\tilde t}_{1}}$, $M_{D_3}$, $\tan \beta$, $\mu$ and $M_2$
define the various scenarios discussed below and are specified in 
\tab{tab:scen}. 
 We use the GUT inspired relation $M_1 = 5 \tan^2
\theta_W M_2 /3$ to reduce the number of free parameters.
 We will mainly
discuss the dependence on the stop mixing angle as this is the most
interesting parameter giving rise to the strongest dependence.
The
parameter $A_t$ can be calculated from $m_{{\tilde t}_{1}}$ and $\cost$
and we have explicitly checked
in all examples that $|A_t| \le 1$~TeV avoiding problems with possible
colour breaking minima. 

\subsection{The MSSM without flavour violation}

\begin{table}[b]
\label{tab:scen}
\caption{Parameters for various scenarios. In all cases we have taken
  the squark mass parameters of the first two generations equal to
  600~GeV, $A_b = - 865$~GeV, $m_{{\tilde t}_2} = 660$~GeV
   and $\mg = 500$~GeV. We use the GUT relation
  for $M_1 = 5 \tan^2 \theta_W M_2 /3$.}
\begin{center}
\begin{tabular}{|c|cccccc|}
\hline 
scenario & $\mst$~[GeV] &  $M_{D_3}$~[GeV] &
           $\tan \beta$ & $M_2$~[GeV] & $\mu$~[GeV] & $m_{A^0}$~[TeV]\\ \hline
1 & 340 & 580 & 6 & 150 & 500 & 1.4\\ \hline
2 & 380 & 580 & 6 & 150 & 500 & 1.4 \\ \hline
3 & 340 & 580 & 30 & 450 & 150 & 0.86\\ \hline
4 & 340 & 550 & 6 & 450 & 150 & 1.4 \\ \hline 
5 & 340 & 550 & 6 & 1100 & 600 & 1.8  \\ \hline 
6 & 340 & 580 & 30 & 1100 & 600 & 0.86  \\ \hline 
\end{tabular}
\end{center}
\end{table}

\begin{figure}[t]
\setlength{\unitlength}{1mm}
\begin{center}
\begin{picture}(83,60)
\put(0,-10){\mbox{\epsfig{figure=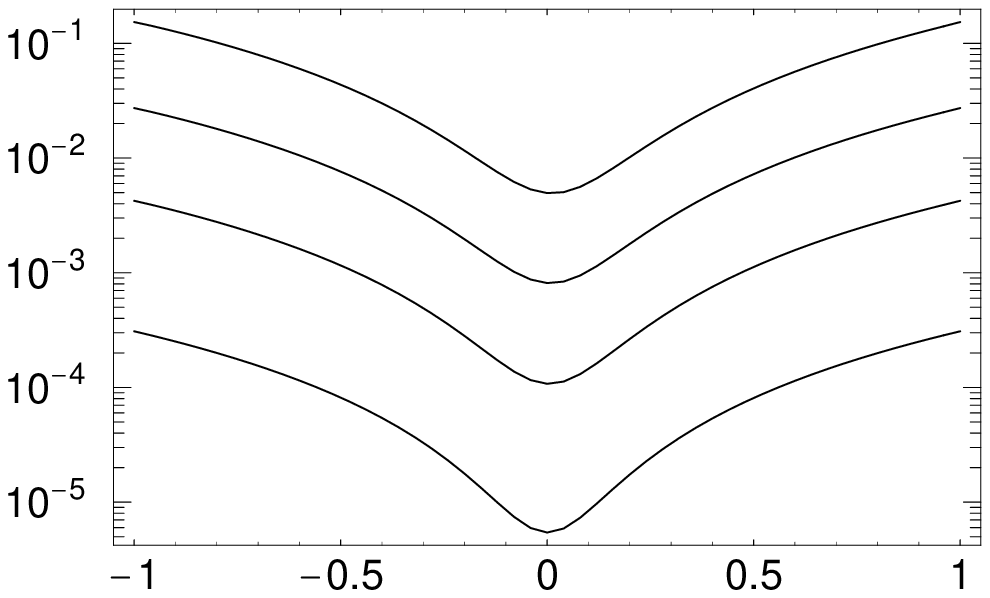,height=7.cm,width=8.cm}}}
\put(0,53){\makebox(0,0)[bl]{{
   $\Gamma(\tilde g \to {\tilde t}_1 W^-  \bar{b})$~[GeV]}}}
\put(83,-5){\makebox(0,0)[br]{{$\cost$}}}
\put(32,47){\makebox(0,0)[br]{{\small 340}}}
\put(32,39){\makebox(0,0)[br]{{\small 360}}}
\put(32,31){\makebox(0,0)[br]{{\small 380}}}
\put(32,18){\makebox(0,0)[br]{{\small 400}}}
\end{picture}
\end{center}
\caption[]{Partial decay width $\Gamma(\tilde g \to {\tilde t}_1 W^- \bar{b})$ 
   as a function of $\cost$ for $\mg = 500$~GeV, $\tan \beta = 6$,
   $m_{{\tilde t}_2} = 660$~GeV, $M_{D_3} = 580$~GeV, $A_b = - 865$~GeV,
   $\mst = 340$, 360, 380 and 400~GeV.}
\label{fig:width}
\end{figure}

In this section we discuss the case of the MSSM without any flavour
violation. 
Before starting with numerical details, we want to note that in practice the 
$\Gamma(\tilde g \to {\tilde t}_1 W b)$ is very well approximated 
(within an error of 1\% and below) by
considering solely the top--quark exchange in \eq{eq:width} except for
the parameter region where the lighter sbottom is nearly on--shell.
The reason for this is that angular momentum conservation at the
$W$-$\tilde t$-$\tilde b$ vertex implies that either the $\tilde t$-$\tilde b$ 
subsystem or one of the $W$-boson squark subsystems form a $P$-wave. 
This in turn implies that the sbottom exchange is 
spin suppressed compared to the top--quark exchange.

In \fig{fig:width} we show the
partial width $\Gamma(\tilde g \to {\tilde t}_1 W b)$ as a function 
  of $\cost$ for $\mst = 340$, 360, 380 and 400~GeV. One encounters a
strong dependence on $\cost$ independent of the mass.
 This is due to the fact that the $W$-boson
couples only to left-handed fermions and correspondingly only to
left sfermions. 
In case of $\cost=0$ the lighter stop is a pure 
right state and the decay is only possible due to a ``spin-flip'' of
the exchanged top quark.

\begin{figure}[t]
\setlength{\unitlength}{1mm}
\begin{center}
\begin{picture}(160,64)
\put(0,-4){\mbox{\epsfig{figure=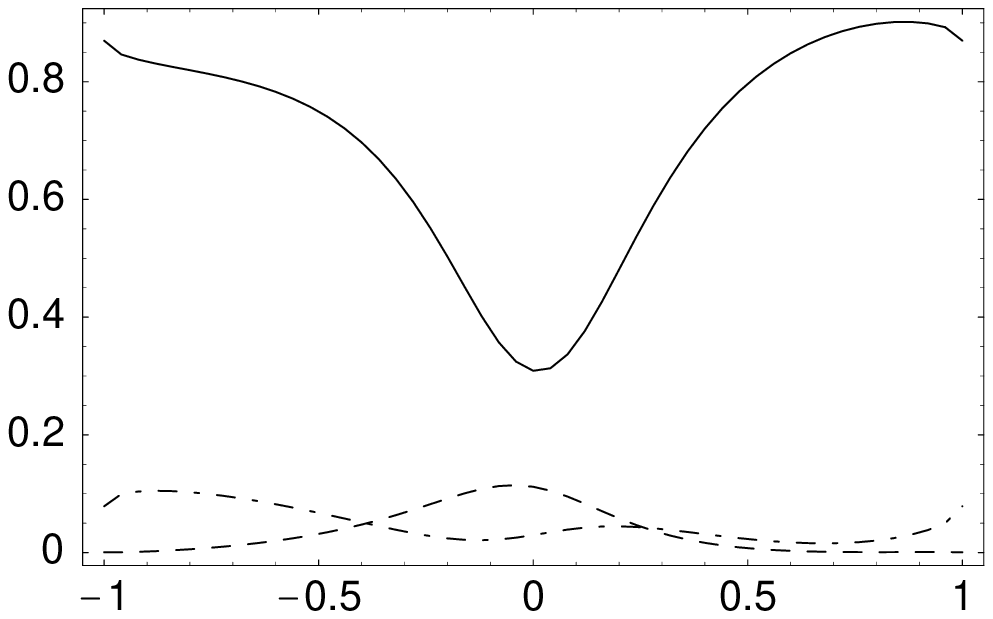,height=7.cm,width=7.cm}}}
\put(-4,61){\makebox(0,0)[bl]{{{\bf a)}}}}
\put(2,60){\makebox(0,0)[bl]{{BR$(\tilde g)$}}} 
\put(70,-1){\makebox(0,0)[br]{{$\cost$}}}
\put(80,-4){\mbox{\epsfig{figure=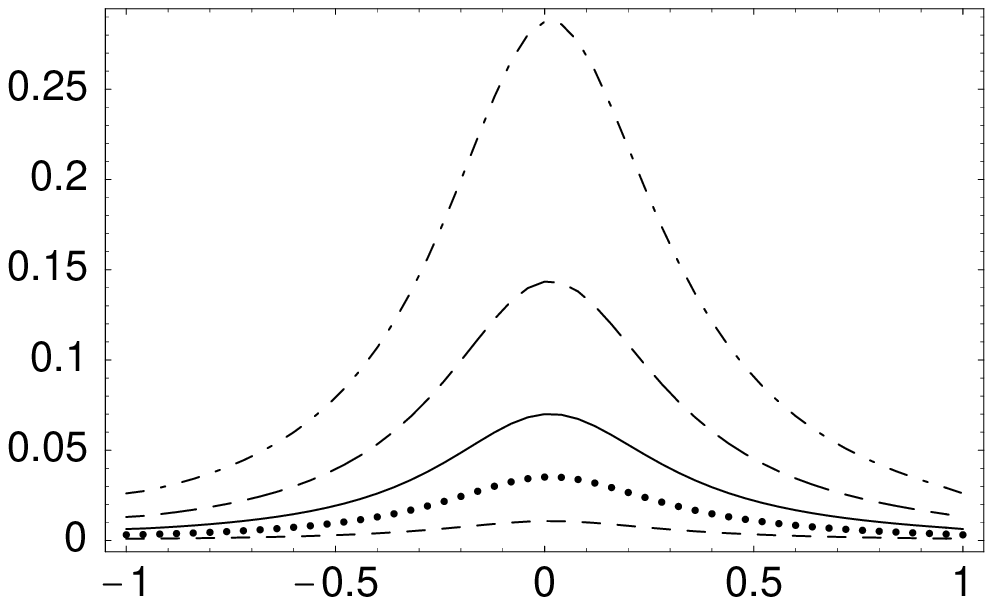,height=7.cm,width=7.cm}}}
\put(77,61){\makebox(0,0)[bl]{{{\bf b)}}}}
\put(83,60){\makebox(0,0)[bl]{{BR$(\tilde g)$}}}
\put(150,-1){\makebox(0,0)[br]{{$\cost$}}}
\end{picture}
\end{center}
\caption[]{Gluino branching ratios for scenario 1 of  \tab{tab:scen}.
           In a) the lines correspond to
 $\tilde g \to \tilde{t}_1 W^- \bar{b} + \bar{\tilde{t}}_1 W^+ b$ (full line),
           $\tilde g \to t \bar{t} \chin{1}$ (dashed line) and
   $\tilde g \to t \bar{b} \chm{1} + \bar{t} b \chp{1}$ (dashed dotted line).
 In b) the lines correspond to 
           $\tilde g \to b \bar{b} \chin{1}$ (dashed line),
           $\tilde g \to b \bar{b} \chin{2}$ (dotted line),
           $\tilde g \to \sum_q q \bar{q} \chin{1}$ (full line),
       $\tilde g \to \sum_q q \bar{q} \chin{2}$ (long short dashed line), and
 $\tilde g \to \sum_{q,q'} q \bar{q}' \chm{1}+ \bar{q} q' \chm{1}$
   (dashed dotted line). 
  $q$ and $q'$ are summed over $u,d,c,s$.}
\label{fig:BrScen1}
\end{figure}

In \figs{fig:BrScen1}{fig:BrScen2} we show the gluino branching ratios as
a function of $\cost$ for $\mst=340$ and $380$ GeV. In the first example
the decay $\tilde g \to \tilde{t}_1 W b$ (full line) dominates for most
of the range. We want to note that this is not a kinematical effect because
$\mst + m_b + m_W \simeq 2 m_t + \mchin{1} > m_t + m_b + \mchpm{1}$.
In the second case the decay $\tilde g \to \tilde{t}_1 W b$ is about as 
important as $\tilde g \to t b \chpm{1}$ despite the fact that it
is kinematically
suppressed compared to the final states with the lighter chargino.
\begin{figure}[h]
\setlength{\unitlength}{1mm}
\begin{center}
\begin{picture}(160,64)
\put(0,-4){\mbox{\epsfig{figure=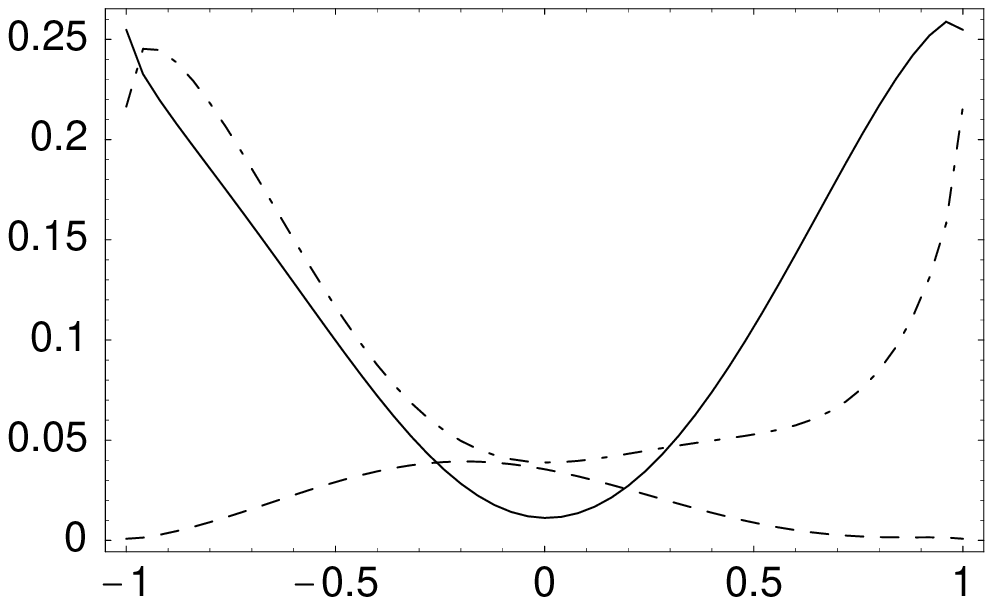,height=7.cm,width=7.cm}}}
\put(-4,61){\makebox(0,0)[bl]{{{\bf a)}}}}
\put(2,60){\makebox(0,0)[bl]{{BR$(\tilde g)$}}} 
\put(70,-1){\makebox(0,0)[br]{{$\cost$}}}
\put(80,-4){\mbox{\epsfig{figure=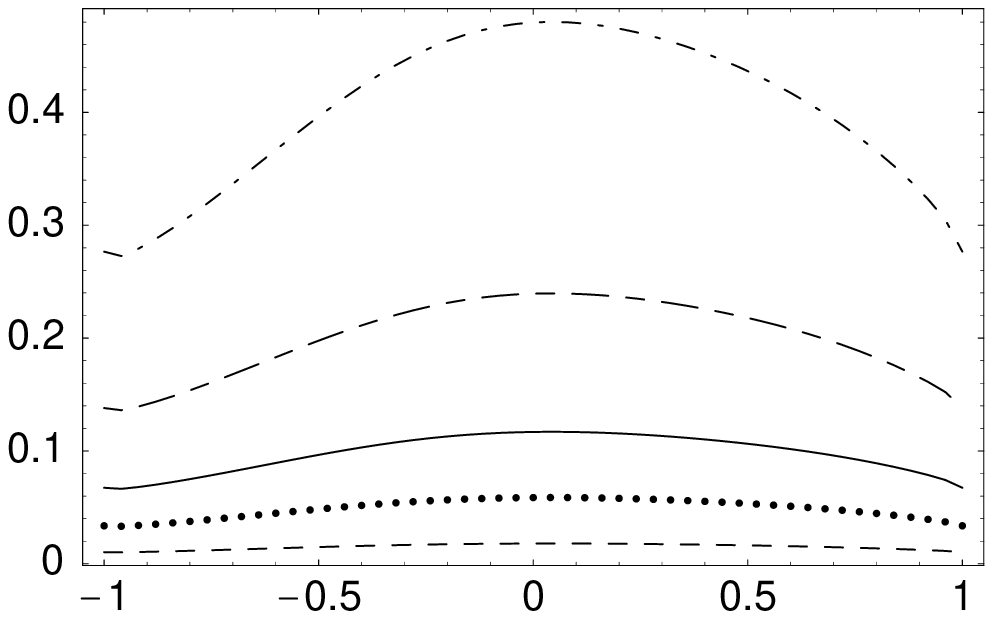,height=7.cm,width=7.cm}}}
\put(77,61){\makebox(0,0)[bl]{{{\bf b)}}}}
\put(83,60){\makebox(0,0)[bl]{{BR$(\tilde g)$}}}
\put(150,-1){\makebox(0,0)[br]{{$\cost$}}}
\end{picture}
\end{center}
\caption[]{Gluino branching ratios for scenario 2. The parameters are
           specified in \tab{tab:scen} and in the text.
          In a) the lines correspond to
 $\tilde g \to \tilde{t}_1 W^- \bar{b} + \bar{\tilde{t}}_1 W^+ b$ (full line),
           $\tilde g \to t \bar{t} \chin{1}$ (dashed line) and
   $\tilde g \to t \bar{b} \chm{1} + \bar{t} b \chp{1}$ (dashed dotted line).
 In b) the lines correspond to 
           $\tilde g \to b \bar{b} \chin{1}$ (dashed line),
           $\tilde g \to b \bar{b} \chin{2}$ (dotted line),
           $\tilde g \to \sum_q q \bar{q} \chin{1}$ (full line),
       $\tilde g \to \sum_q q \bar{q} \chin{2}$ (long short dashed line), and
 $\tilde g \to \sum_{q,q'} q \bar{q}' \chm{1}+ \bar{q} q' \chm{1}$
   (dashed dotted line). 
  $q$ and $q'$ are summed over $u,d,c,s$.}
\label{fig:BrScen2}
\end{figure}
\begin{figure}[ht]
\setlength{\unitlength}{1mm}
\begin{center}
\begin{picture}(160,64)
\put(0,-4){\mbox{\epsfig{figure=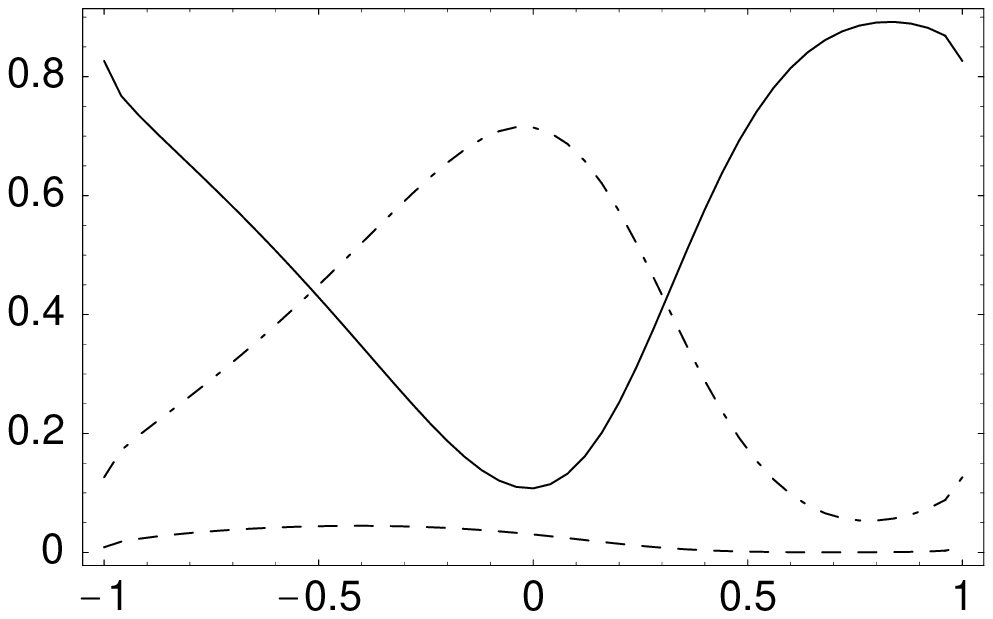,height=7.cm,width=7.cm}}}
\put(-4,61){\makebox(0,0)[bl]{{{\bf a)}}}}
\put(2,60){\makebox(0,0)[bl]{{BR$(\tilde g)$}}} 
\put(70,-1){\makebox(0,0)[br]{{$\cost$}}}
\put(80,-4){\mbox{\epsfig{figure=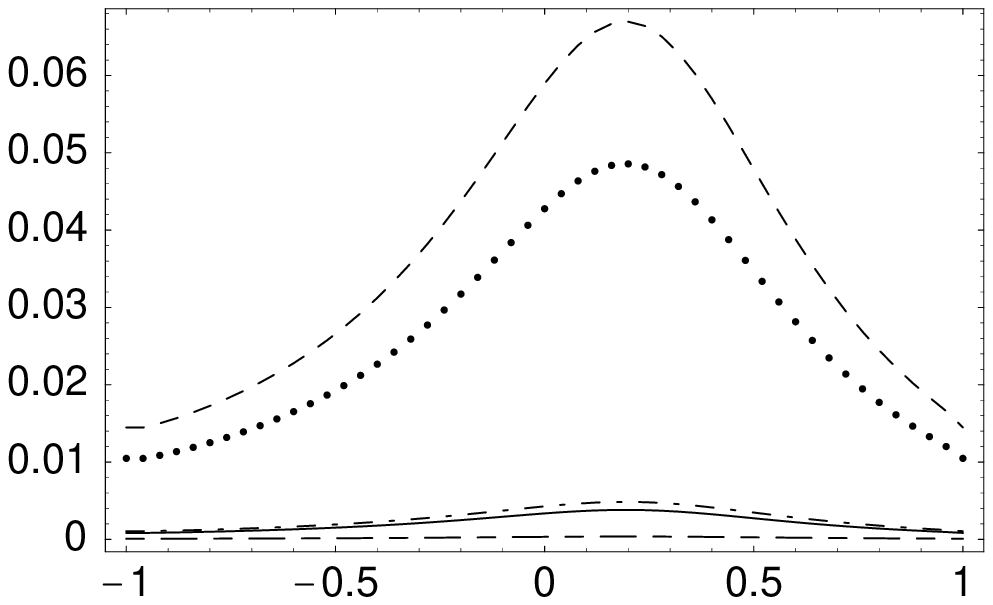,height=7.cm,width=7.cm}}}
\put(77,61){\makebox(0,0)[bl]{{{\bf b)}}}}
\put(83,60){\makebox(0,0)[bl]{{BR$(\tilde g)$}}}
\put(150,-1){\makebox(0,0)[br]{{$\cost$}}}
\end{picture}
\end{center}
\caption[]{Gluino branching ratios for scenario 3. The parameters are
           specified in \tab{tab:scen}. In a) the lines correspond to
 $\tilde g \to \tilde{t}_1 W^- \bar{b} + \bar{\tilde{t}}_1 W^+ b$ (full line),
           $\tilde g \to t \bar{t} \chin{1}$ (dashed line) and
   $\tilde g \to t \bar{b} \chm{1} + \bar{t} b \chp{1}$ (dashed dotted line).
 In b) the lines correspond to 
           $\tilde g \to b \bar{b} \chin{1}$ (dashed line),
           $\tilde g \to b \bar{b} \chin{2}$ (dotted line),
           $\tilde g \to \sum_q q \bar{q} \chin{1}$ (full line),
       $\tilde g \to \sum_q q \bar{q} \chin{2}$ (long short dashed line), and
 $\tilde g \to \sum_{q,q'} q \bar{q}' \chm{1}+ \bar{q} q' \chm{1}$
   (dashed dotted line). 
  $q$ and $q'$ are summed over $u,d,c,s$.}
\label{fig:BrScen5}
\end{figure}
\begin{figure}[ht]
\setlength{\unitlength}{1mm}
\begin{center}
\begin{picture}(160,64)
\put(0,-4){\mbox{\epsfig{figure=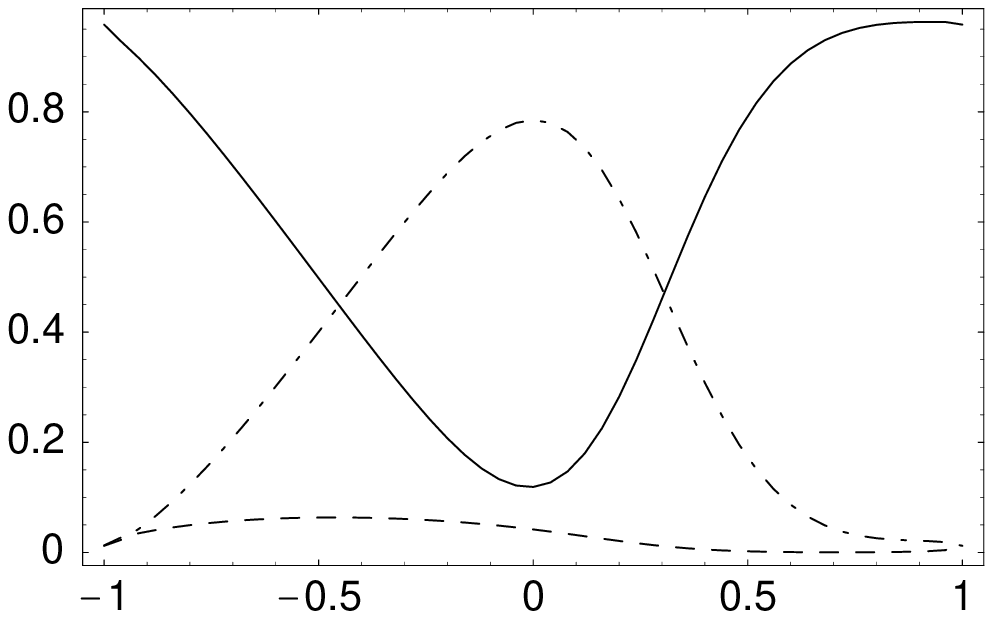,height=7.cm,width=7.cm}}}
\put(-4,61){\makebox(0,0)[bl]{{{\bf a)}}}}
\put(2,60){\makebox(0,0)[bl]{{BR$(\tilde g)$}}} 
\put(70,-1){\makebox(0,0)[br]{{$\cost$}}}
\put(80,-4){\mbox{\epsfig{figure=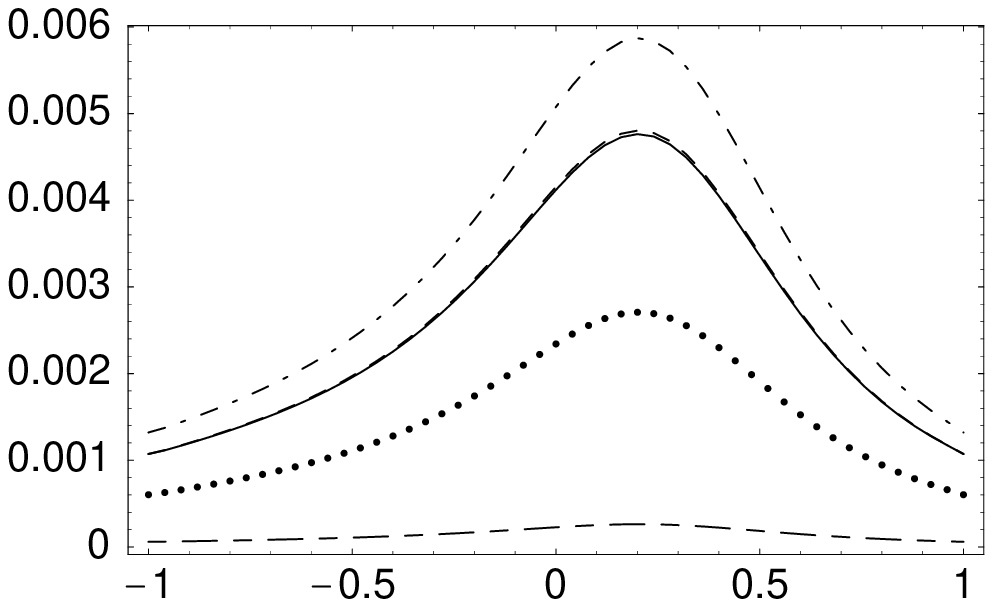,height=7.cm,width=7.cm}}}
\put(77,61){\makebox(0,0)[bl]{{{\bf b)}}}}
\put(83,60){\makebox(0,0)[bl]{{BR$(\tilde g)$}}}
\put(150,-1){\makebox(0,0)[br]{{$\cost$}}}
\end{picture}
\end{center}
\caption[]{Gluino branching ratios for scenario 4. The parameters are
           specified in \tab{tab:scen}. In a) the lines correspond to
 $\tilde g \to \tilde{t}_1 W^- \bar{b} + \bar{\tilde{t}}_1 W^+ b$ (full line),
           $\tilde g \to t \bar{t} \chin{1}$ (dashed line) and
   $\tilde g \to t \bar{b} \chm{1} + \bar{t} b \chp{1}$ (dashed dotted line).
 In b) the lines correspond to 
           $\tilde g \to b \bar{b} \chin{1}$ (dashed line),
           $\tilde g \to b \bar{b} \chin{2}$ (dotted line),
           $\tilde g \to \sum_q q \bar{q} \chin{1}$ (full line),
       $\tilde g \to \sum_q q \bar{q} \chin{2}$ (long short dashed line), and
 $\tilde g \to \sum_{q,q'} q \bar{q}' \chm{1}+ \bar{q} q' \chm{1}$
   (dashed dotted line). 
  $q$ and $q'$ are summed over $u,d,c,s$.}
\label{fig:BrScen6}
\end{figure}
In both cases the asymmetry in $\cost$ is mainly due to interference
effects in the width for $\tilde g \to t b \chpm{1}$ between stop and
sbottom contributions. Together with information from stop production and
decays \cite{StopZwei} this asymmetry can be used to get information on the
sign of $\cost$. Note that this is asymmetry is not physical in the sense
that it can be measured but is given by the structure of the theory
implying that the experimental findings together with the consistency
of the theory tells one which sign of $\cost$ is realized in nature.
In addition we want to note that the branching ratio for the
final state $t \bar{t} \chin{1}$ is maximal near $\cost=0$ because in this
case the right stop contributes most.
We have checked that the behaviour shown hardly depends
on $\tan \beta$. 

The situation is somewhat different in scenarios
where the $|\mu| \ll |M_{1,2}$ as can be seen in
\figs{fig:BrScen5}{fig:BrScen6}. Independent of $\tan \beta$ the
channels $\tilde g \to \tilde{t}_1 W b$ (full line)
and $\tilde g \to t b \chpm{1}$ (dashed dotted) line show a strong
dependence on $\cost$. The reason is that the $W$-boson couples the left
states whereas the  higgsino--like chargino couples mainly to the right stop
giving rise to the peak of $\tilde g \to t b \chpm{1}$ near
$\cost=0$. The asymmetry in the sign of $\cost$ can be traced back
to stop-chargino-bottom coupling which reads as 
$ - g R^{{\tilde t}*}_{11} V_{j1} + Y_t R^{{\tilde t} *}_{12} V_{j1}$.
Here the relative sign (in case of complex parameters relative phase) 
between $R^{\tilde t}_{11}$ and $R^{\tilde t}_{12}$ gives rise to the 
asymmetry. In both cases the final state $t \bar{t} \chin{1}$ is less
important because the lightest neutralino is to a large extent a down-type
higgsino and the decay into the second lightest neutralino, which is
mainly a up-type higgsino, is kinematically suppressed. This ordering
is beside kinematics also the reason for 
BR$(\tilde  g \to b \bar{b} \chin{1}) > $  
Br$(\tilde g \to b \bar{b} \chin{2})$ in these scenarios. We want to
stress here that for small $\tan \beta$ the gluino mainly decays
into modes containing at least one top or a stop in higgsino--like scenarios
as can be seen in  \fig{fig:BrScen6}.

\begin{figure}[t]
\setlength{\unitlength}{1mm}
\begin{center}
\begin{picture}(83,60)
\put(0,-10){\mbox{\epsfig{figure=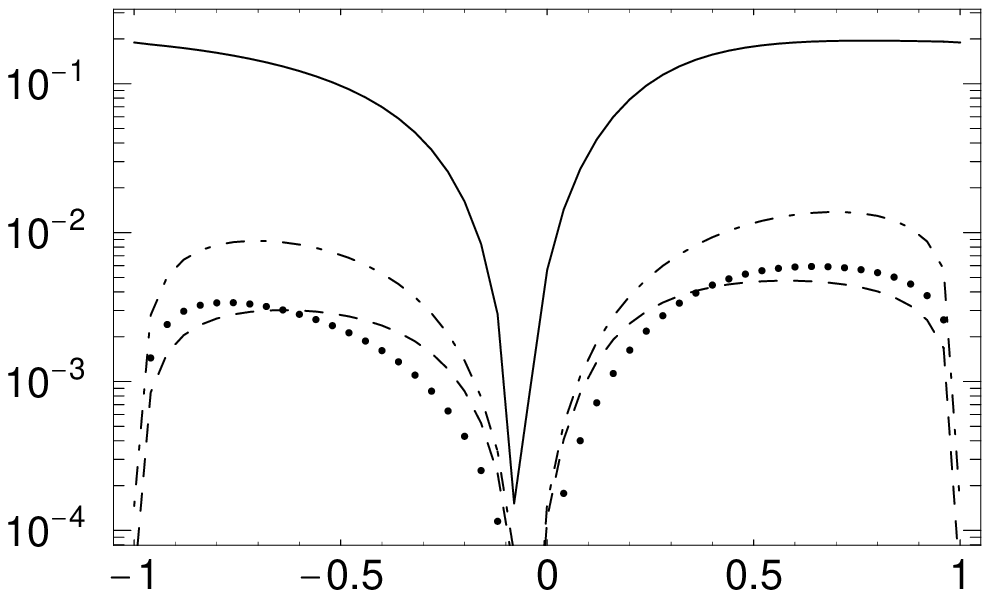,height=7.cm,width=8.cm}}}
\put(0,53){\makebox(0,0)[bl]{{
  BR$(\tilde g \to {\tilde t}_1 \bar{c}+ {\tilde t}_1^* c)$}}}
\put(83,-5){\makebox(0,0)[br]{{$\cost$}}}
\end{picture}
\end{center}
\caption[]{Branching ratio 
    $\tilde g \to {\tilde t}_1 \bar{c}+ \bar{\tilde t}_1 c$ 
   as a function of $\cost$ for scenarios 1 (dashed line),
   2 (dashed dotted line), 3 (full line) and 4 (dotted line).
   The parameters are given in \tab{tab:scen} and in the text.}
\label{fig:BrStC}
\end{figure}

The signature of $\tilde g \to \tilde{t}_1 W b$ clearly depends on the
decay modes of the lighter stop. In the examples studied here ${\tilde t}_1$ 
decays mainly into $b {\tilde \chi}^+_1$ and the chargino
decays further mainly into $\chin{1} q' \bar{q}$ and $\chin{1} l \nu$.
Depending on the parameters chosen, other 2-body decay modes of
$\tilde{t}_1$ can become important \cite{StopZwei} or higher order
decay modes are important in case all the two-body tree-level decay modes are
kinematically  forbidden \cite{Hikasa:1987db,StopDrei}.  
If for example
the decay $\tilde t_1$ mode is $\tilde t_1\to b W \chin{1}$ one gets a
final state $\tilde g \to W^+ W^- b \bar{b} \chin{1}$. The same final
state can be obtained via the chain $\tilde g \to t \bar{t} \chin{1}
\to W^+ W^- b \bar{b} \chin{1}$. Similarly, one can show that
for all stop final states one finds a gluino decay into a neutralino or
a chargino that contains the same particles in the final state.
However, in general the energy
distribution as well as the angular distribution 
of the final state particles will be different, which is
of course important for gluino searches and measurement of gluino properties.

\subsection{The MSSM with flavour violation}

Let us now turn to the case that case that the flavour violating
coupling gluino -- stop -- charm-quark is non-zero. As mentioned in
\sect{sec:formula} we use the formulas given in \cite{Hikasa:1987db}
to describe the mixing between top-squarks and the left scalar charm.
In \fig{fig:BrStC} we
display the branching ratios for the scenarios 1 -- 4 of \tab{tab:scen}.
We have checked that the used values for the stop -- scalar charm mixing
are compatible with the bounds given in \cite{Gabbiani:1996hi}.
 One clearly sees that under the
assumption of minimal flavour violation for small 
$\tan \beta$ the branching ratio is at most 1 -- 2 \%.
However, in case of large $\tan \beta$ this decay mode can be potentially
large giving branching ratios up to 20\%. In case that this decay
is important the main consequence is a reduction in the multiplicity
of the final state compared to the other gluino decays. 

As mentioned in \sect{sec:formula} the used approximation for the description
of the scalar-charm -- stop mixing is the result of
a single step integration of the corresponding RGEs assuming the
CKM matrix is the only source of flavour violation at the high scale.
Therefore, the results obtained above should not be taken literally but
as a demonstration of the expected order of magnitude under the
assumptions above. Clearly, additional flavour violation in the squark
sector at the high scale could enlarge the branching ratio for the decay
$\tilde g \to {\tilde t}_1 \bar{c}$. However, it cannot be much larger than
20\%  because otherwise the experimental bounds 
from low energy physics \cite{Gabbiani:1996hi} will be violated. Moreover,
in case of additional flavour violation in the squark
sector also the $\tilde g \to {\tilde t}_1 \bar{u}$ could be 
potentially large because there is not necessarily a suppression of the form
$| V_{ub}/ V_{cb}|^2$ in such a case.

\subsection{R-parity violation}

\begin{figure}[t]
\setlength{\unitlength}{1mm}
\begin{center}
\begin{picture}(83,60)
\put(0,-10){\mbox{\epsfig{figure=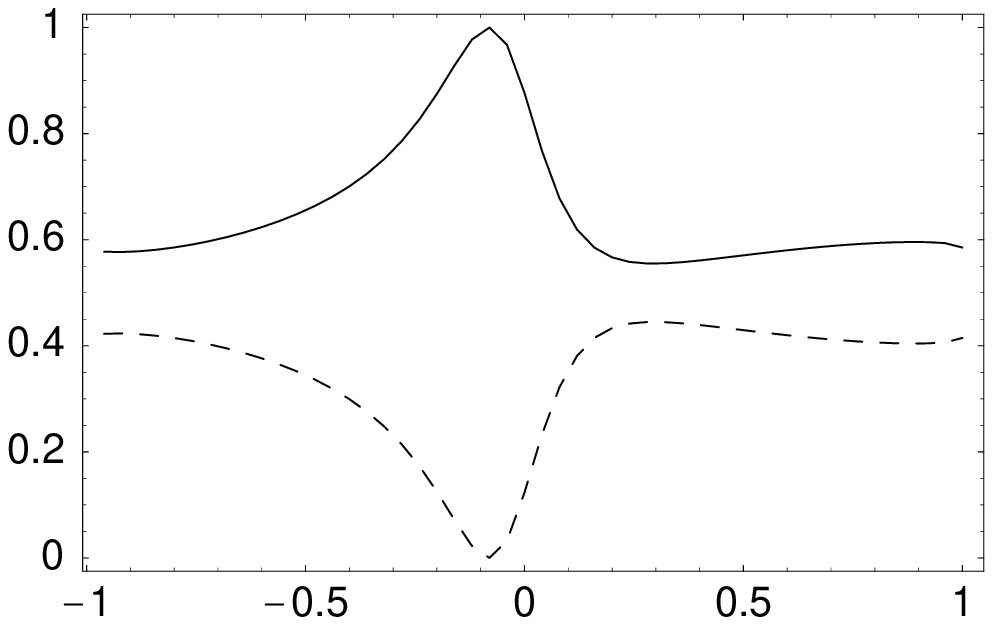,height=7.cm,width=7.5cm}}}
\put(4,53){\makebox(0,0)[bl]{{BR$(\tilde g)$}}}
\put(78,-5){\makebox(0,0)[br]{{$\cost$}}}
\end{picture}
\end{center}
\caption[]{Gluino branching ratio for scenarios 6
   of \tab{tab:scen}. The lines correspond to
   $\tilde g \to {\tilde t}_1 W^- \bar{b} + \bar{{\tilde t}}_1 W^+ b$ 
   (full line) and
   $\tilde g \to {\tilde t}_1 \bar{c} + \bar{{\tilde t}}_1 c$ (dashed line).}
\label{fig:BrWBStRp}
\end{figure}
In this section we are going to study scenarios where R-parity is
broken. Here we focus on the model where one adds bilinear terms
to the superpotential as well as to the soft SUSY breaking part
\cite{Diaz:1997xc}. This class of models can successfully explain
the neutrino mass hierarchy as well as the neutrino mixing angles
(see \cite{NuPaper} and references therein). It also resembles
in many respects the models with explicit lepton number violating trilinear
couplings $\lambda$ and $\lambda'$ \cite{Dreiner:wm}. This can easily be
seen by a rotation of the superfields where one transforms the bilinear
terms in the superpotential into trilinear 
terms\footnote{This still leaves bilinear terms in the soft SUSY breaking
part of the Lagrangian.} leading to
$\lambda_{ijk} \propto \epsilon_i Y^E_{jk}/\mu$ and 
$\lambda_{ijk}' \propto \epsilon_i Y^D_{jk}/\mu$ \cite{Ferrandis:1998ii}. Here
$Y^E$ and $Y^D$ are the lepton and down--quark Yukawa couplings and
$\epsilon_i$ are the parameters of the bilinear terms violating
lepton number in the superpotential.

In what follows we study the scenarios discussed above as well as
scenarios where the stop is the LSP and the gluino is the next heavier
supersymmetric particle.
The latter two scenarios can e.g.~be obtained
in GMSB scenarios \cite{Raby:1997bp} or in string scenarios 
 \cite{Kaplunovsky:1993rd}.
In the bilinear
model the data from neutrino experiments imply relatively small
R-parity violating couplings, 
e.g.~$|\epsilon_i /\mu |\simeq O(10^{-3})$ \cite{NuPaper}
(for a discussion on gluino decays  and production in the case
of larger R-parity violating parameters see \cite{GluRP,Xi:2001cy}
and references therein). This implies
that branching ratios for the gluino decays in scenarios 1 -- 4
are practically the same as in the R-parity conserving case because
the R-parity violating decay modes have a branching ratio
of at most O($10^{-7}$). However,
now the lightest neutralino will decay further giving rise
to additional jets and leptons in the final state \cite{NeutRP1,NeutRP1b} 
compared to the R-parity conserving case (for
recent discussions of neutralino decays with trilinear R-parity
couplings see e.g.~\cite{NeutRP2}).

In scenarios 5 and 6 of \tab{tab:scen}
 the stop is the LSP and the 
gluino  is the next to lightest SUSY particle. Scenario 5
is a low $\tan \beta$ scenario whereas scenario 6 is
 a large $\tan \beta$ scenario. In the case of
small $\tan \beta$ 
the three--body decay dominates practically with 100\%. Moreover, the
stop  decays mainly into a lepton and a $b$-quark
\cite{StopRp}. Thus, the signature is in this case 2 $b$-jets,
a $W$-boson and a charged lepton. In the case of large $\tan \beta$ also
the decay into ${\tilde t}_1 \bar{c}$ becomes important as can be seen in
\fig{fig:BrWBStRp}. The corresponding signature is in this case a
$b$-jet, a $c$-jet and a charged lepton. Note, that due to
the Majoranna nature of the gluino the final state leptons can have
same sign which clearly reduces the Standard Model background.

Let us finally comment on the case of gluino LSP. It decays in these
scenarios into the following final states: $q \bar{q} \nu_i$ and $q'
\bar{q} l^\pm$ and $g \nu_i$. For the scenarios discussed above the
width of the gluino varies between O(eV) and  O(keV). It turns out
that in such a case final states containing  top-quarks and/or
bottom quarks dominate and that the branching ratios are
sensitive to neutrino mixing angles similar to the case of neutralinos
\cite{NeutRP1b}. This, however, is beyond the scope of this paper
and will be discussed in a dedicated paper \cite{GluinoNeu}.

\section{Conclusions}
\label{sec:con}

We have computed and studied the decays 
$\tilde g \to {\tilde t}_1 W^- \bar{b}$ and
$\tilde g \to {\tilde t}_1 \bar{c}$. We have demonstrated that the
branching ratio of $\tilde g \to {\tilde t}_1 W^- \bar{b}$ is large
and can even be dominant. This decay does not lead to new final states
compared to the decays $\tilde g \to t \bar{t} \chin{i}$ or
$\tilde g \to t b \chpm{j}$. However, it  leads in general to different
energy and angular distributions of the final state particle
compared to the decays into a chargino or a neutralino. 
This is clearly of importance
for searches of gluinos at present and future colliders. 
In addition we have worked out possible signatures in models where R-parity is
broken by  lepton number violating terms.

\acknowledgments
This work is supported by the Erwin Schr\"odinger fellowship Nr.~J2095
of the `Fonds zur F\"orderung der wissenschaftlichen Forschung' of
Austria and partly by the `Schweizer Nationalfonds'.

\end{document}